\documentclass[a4paper,fleqn,usenatbib]{mnras}

\usepackage{graphicx, rotate}
\usepackage{natbib, lscape}
\usepackage{latexsym, amssymb, verbatim}
\usepackage{eufrak}
\usepackage{amsmath}
\usepackage{epsfig}
\usepackage{times}
\usepackage{rotating}
\usepackage{graphicx}
\usepackage{color}
\usepackage{multirow}

\newcounter{chem}
\newcounter{temp}

\newcommand{\CII}{[C{\sc ii}]} 
\newcommand{\kms}{km~s$^{-1}$}
\newcommand{\vlsr}{$v_{\rm lsr}$}

\title[The Inception of Star Cluster Formation]{The Inception of Star Cluster Formation Revealed by \CII\ Emission Around an Infrared Dark Cloud}
\author[Bisbas, T.~G.]{
Thomas G. Bisbas,$^{1,2}$\thanks{E-mail: tbisbas@virginia.edu}
Jonathan C. Tan,$^{1,3}$
Timea Csengeri,$^{4}$
Benjamin Wu,$^{5}$\newauthor
Wanggi Lim,$^{6}$
Paola Caselli,$^{2}$
Rolf G\"usten,$^{4}$
Oliver Ricken,$^{4}$
and Denise Riquelme$^{4}$
\\
$^{1}$Department of Astronomy, University of Virginia, Charlottesville, VA 22904, USA\\
$^{2}$Max-Planck-Institut f\"ur Extraterrestrische Physik, Giessenbachstrasse 1, D-85748 Garching, Germany\\
$^{3}$Department of Space, Earth \& Environment, Chalmers University of Technology, Gothenburg, Sweden\\
$^{4}$Max-Planck-Institut f\"ur Radioastronomie, auf dem H\"ugel 69, 53121, Bonn, Germany\\
$^{5}$National Astronomical Observatory of Japan, Mitaka, Tokyo 181-8588, Japan\\
$^{6}$SOFIA-USRA, NASA Ames Research Center, MS 232-12, Moffett Field, CA 94035, USA
}

\date{Accepted XXX. Received YYY; in original form ZZZ}

\pubyear{2015}

\begin{document}
\label{firstpage}
\pagerange{\pageref{firstpage}--\pageref{lastpage}}
\maketitle

\begin{abstract}
We present {\it SOFIA-upGREAT} observations of \CII\ emission of
Infrared Dark Cloud (IRDC) G035.39-00.33, designed to trace its atomic
gas envelope and thus test models of the origins of such clouds. 
Several velocity components of \CII\ emission are detected,
tracing structures that are at a wide range of distances in the
Galactic plane. We find a main component that is
  likely associated with the IRDC and its immediate surroundings.
This strongest emission component has a velocity similar to that of
the $^{13}$CO(2-1) emission of the IRDC, but offset by $\sim3\:{\rm
  km\:s}^{-1}$ and with a larger velocity width of $\sim9\:{\rm
  km\:s}^{-1}$. The spatial distribution of the \CII\ emission of this
component is also offset predominantly to one side of the dense
filamentary structure of the IRDC.
The C{\sc ii} column density is estimated to be of the order of $\sim10^{17}-10^{18}\,{\rm cm}^{-2}$.
We compare these results to the \CII\ emission from numerical
simulations of magnetized, dense gas filaments formed from giant
molecular cloud (GMC) collisions, finding similar spatial and
kinematic offsets.
These observations and modeling of \CII\ add further to the evidence
that IRDC G035.39-00.33 has been formed by a process of GMC-GMC
collision, which may thus be an important mechanism for initiating star cluster formation.
\end{abstract}

\begin{keywords}
{ISM: kinematics and dynamics --- ISM: clouds --- radiative transfer}
\end{keywords}

%%%%%%%%%%%%%%%%%%%%%%%
\section{Introduction}%
%%%%%%%%%%%%%%%%%%%%%%%

Infrared Dark Clouds (IRDCs) are dense parts of molecular clouds,
discovered due to their mid-infrared (MIR) absorption of Galactic
background light \citep{Pera96, Egan98}. Their characteristic masses,
sizes and densities, indicate that they are likely to be the
precursors of star-forming clumps and star clusters
\citep[e.g.,][]{Tan14}. Since most stars form in such clusters, the
study of IRDCs can thus reveal crucial information on the processes
that set global galactic star formation rates.
There have been many observational studies of IRDCs \citep[see review
  by, e.g.,][]{Tan14}. One well-studied sample of IRDCs is that
selected by \citet{Butl09,Butl12}, i.e., ten IRDCs (named A-J) for
which methods of MIR extinction mapping to measure cloud structure
were developed. \citet{Kain13} then developed combined near-infrared
(NIR) and MIR extinction maps, which more accurately probe the lower
column density envelopes of these clouds.
%\citet{Lim14} studied far-infrared (FIR) extinction mapping and
%searched for extinction law variations that may indicate dust grain
%evolution.
\citet{Hern15} studied the kinematics and dynamics of these
ten IRDCs and their surrounding giant molecular clouds (GMCs) using
$^{13}$CO Galactic Ring Survey (GRS) data \citep{Jack2006}.

From the above sample of ten IRDCs, G035.39-00.33 (IRDC H) was mapped
in many molecular tracers with the IRAM 30m telescope. This IRDC is a
highly filamentary cloud at a near kinematic distance of $2.9\,{\rm
  kpc}$ \citep{Simo06}. These observations detected parsec scale
SiO(2-1) emission, which is potentially caused by large-scale shocks
due to converging flows and/or a collision of two molecular clouds
\citep{Jime10}. These data were also used to discover widespread CO
depletion in the dense IRDC filament \citep{Hern11}, as well as
constraining filament dynamics \citep{Hern12}, with indications that
its inner regions are virialized. \citet{Hens13} found velocity
offsets of dense N$_2$H$^+$ emitting gas compared to lower density
C$^{18}$O emitting gas. With IRAM PdBI, \citet{Hens14} identified
several sub-filaments within IRDC~H, analogous to the structures
discovered by \citet{Haca13} in Taurus. Other studies of this IRDC
have included measurement of widespread deuteration \citep{Barn16} and
its temperature structure \citep{Soko17}.

To date there have been relatively few previous studies of \CII\ from IRDCs. \citet{Beut14} studied the carbon cycle (CO, C{\sc i}, C{\sc  ii}) in very localized regions (typically just a few parsecs in size) within several IRDCs using {\it Herschel-HIFI} for three clouds and {\it SOFIA-GREAT} for one cloud. \CII\ was detected in three regions; the mass of the gas in C{\sc ii} was always higher by a factor of a few than that of C{\sc i}. Averaging over these regions, the \CII\ lines were observed to be relatively broad ($\gtrsim10\,{\rm km}\,{\rm s}^{-1}$) and with integrated intensities of several ${\rm K}\,{\rm km}/{\rm s}$. While the [C{\sc ii}] emission depends on the strength of the radiation field, they also found additional signatures of multiple velocity components and strong velocity gradients that indicate that other processes, e.g., energetic gas flows, may also be contributing to the \CII\ excitation and perhaps be involved in the formation of IRDCs.

In this {\it Letter}, we present pilot \CII\ {\it SOFIA-upGREAT} observations of IRDC~H, with the primary goal of understanding the lower density atomic layers around the dense gas structure to constrain models of the formation conditions of such clouds and thus the initiation of star cluster formation.

%%%%%%%%%%%%%%%%%%%%%%%
\section{Observations}%
\label{sec:obs}       %
%%%%%%%%%%%%%%%%%%%%%%%

We obtained a fully sampled map of the 158~$\rm \mu m$ atomic fine
structure line of \CII\ with the upGREAT receiver \citep{Risa16}
on-board {\it SOFIA} on 18th May 2016 as part of the program 04\_0169
(PI: Tan). %We describe these as ``pilot'' observations of IRDCs for this project, since the amount of observing time obtained was   $\sim$80  minutes, including overheads.  This limited the achieved sensitivity and the extent of the region that could be mapped to a relatively small area around the IRDC.
The upGREAT array was tuned to the \CII\ line at $1900.537\:\rm GHz$
in the lower sideband using both the H and V polarizations and the OTF mode.  The
low-frequency channel was tuned to the $^{12}$CO(11-10) line at
$1267.0\:\rm GHz$ in the lower sideband of the receiver.
The observations were carried out at an altitude of 43\,kft for a
total on-source integration time of $\sim$33 minutes and with an
atmospheric water vapor content below 10\,$\mu$m, resulting in an
atmospheric opacity of $0.18-0.22$ at the frequency of the
\CII\ line. An OFF position of $l=35.6987^{\circ}$, $b=-0.6019^{\circ}$ 
was used.
The data were calibrated with the KOSMA/GREAT calibrator
\citep{Guan2012}, and further processed with the CLASS software.  The
overall absolute flux calibration is expected to be within the nominal
10\% uncertainty.

The spectra were converted to the $T_{\rm mb}$ temperature scale with
a main beam efficiency determined for each pixel
based on observations of Jupiter. The average
efficiency is $\langle\eta^{\rm LFA}_{\rm mb}\rangle=0.65$ for the upGREAT array,
and $\langle\eta^{\rm L1}_{\rm mb}\rangle=0.66$ for the L1 channel.  
We then processed the $^{12}$CO(11-10) and the \CII\ data in the same way. 
A third order baseline was subtracted from each spectrum, excluding the
velocity range of the \CII\ emission averaged over the map.  The beam
is 17.3\arcsec\ for the $^{12}$CO(11-10), and 14.8\arcsec\ for the
\CII\ observations, respectively, and the spectra were gridded using a
sampling of 1/3 of the beam for both lines.  The maps extend over an
area of $360\arcsec \times 520\arcsec$, covering the main part of the
IRDC filament as well as its immediate surroundings.  We measure a
noise level of 0.5\,K in a single beam for the \CII\ line in
0.3\,\kms\ velocity bins, and 1.04\,K for the $^{12}$CO(11-10) line in the
lowest noise part of the spectrum. The $^{12}$CO(11-10) line is not detected,
therefore in the following we focus on the \CII\ data.

\section{Results}\label{sec:results}

\begin{figure}
\center
\includegraphics[width=0.4\textwidth]{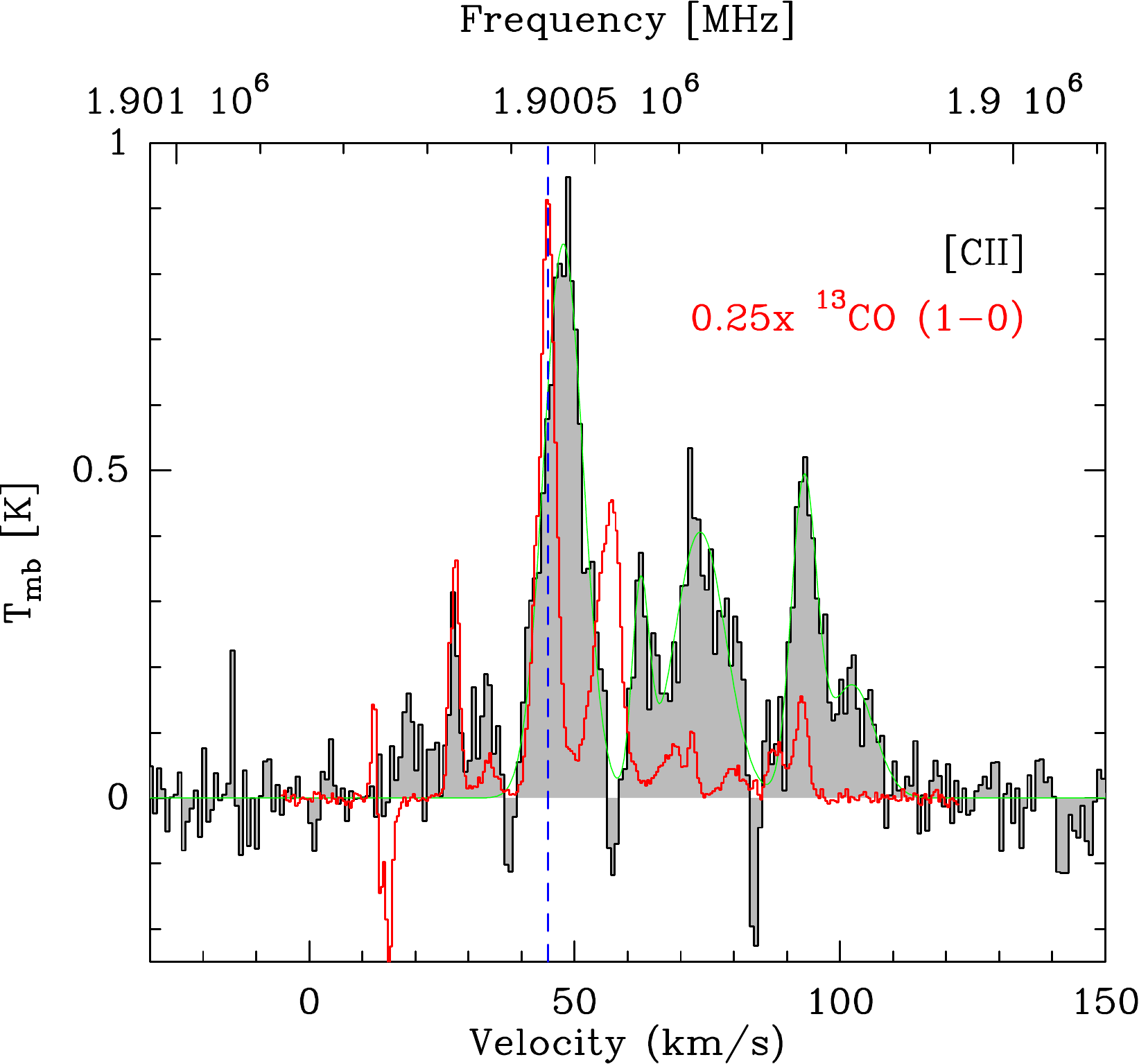}
\caption{
Gray filled histogram shows the [CII] spectrum averaged over the
entire mapped region of IRDC H. The GRS $^{13}$CO(1-0) spectrum
averaged over the same region is shown in red. The upper x axis shows
the signal band frequency scale for the [CII] data. For better
visibility, the $^{13}$CO spectrum is scaled by a factor of 0.25. The
blue dashed line shows $v_{\rm lsr}$ of the IRDC. 
}\label{fig:average_spec}
\end{figure}

In Fig.\,\ref{fig:average_spec} we show the \CII\ spectrum averaged
over the observed region. Several emission peaks are seen. The
strongest has a \vlsr of about 48~$\rm km\:s^{-1}$. Other emission is
seen over a broad velocity range, including weak features at
$\sim30\:{\rm km\:s}^{-1}$ and relatively strong features up to
$\sim110\:{\rm km\:s}^{-1}$. We note that the observed negative
features, e.g., near 70~$\rm km\:s^{-1}$ and possibly near 38 and
57$\:{\rm km\:s}^{-1}$, are likely from contamination of
\CII\ emission in the OFF position.  We fit five Gaussian components
to the spectrum and list the results in Table~\ref{table:3}.
Figure\,\ref{fig:average_spec} also shows the $^{13}$CO(1-0) spectrum
obtained by the GRS survey \citep{Jack2006} over the same region. The
\vlsr\ of the IRDC is at $\sim45\:{\rm km\:s}^{-1}$ as determined from
this $^{13}$CO(1-0) emission \citep[e.g.,][]{Hern15}, and this is
within about 3~$\rm km\:s^{-1}$ of the brightest \CII\ emission
feature. In the velocity range \vlsr$\sim39.7-49.9\:{\rm
    km\:s}^{-1}$, \citet{Hern15} find that the highest peaks of
  brightness temperature of $^{13}$CO(1-0) are associated with IRDC~H,
  however there is also significant contribution from the surrounding
  GMC. For the velocity range $49.9-62.6\,{\rm km}\,{\rm s}^{-1}$,
  which corresponds to the second highest peak in $^{13}$CO emission
  in Fig.~\ref{fig:average_spec}, they find an extended GMC-scale
  component that is connected in position-velocity space to the
  $45\:{\rm km\:s}^{-1}$ feature, of which a significant part overlaps
  on the same line-of-sight as IRDC~H. This indicates that this
  filamentary IRDC may have been formed as a result of an ongoing
  cloud-cloud collision.

\begin{table}
\caption{
Parameters for fitted Gaussian components to the {\CII} spectrum shown
in Fig.~\ref{fig:average_spec}. Columns from left to
right: velocity along the line-of-sight ($v_{\rm lsr}$), velocity
width ($\Delta v$), peak brightness temperature ($T_{\rm mb}$), and
velocity integrated $T_{\rm mb}$ ($\int T_{\rm mb} dv$).}
\label{table:3}      
\centering          
\begin{tabular}{r r c c}
\hline\hline       
$v_{\rm lsr}$ [km/s] & $\Delta v $  [km/s]  &  $T_{\rm mb}$ [K] & $\int T_{\rm mb}$ dv [K km/s]  \\ 
\hline                    
 $47.87\pm  0.16$&  $8.76\pm  0.36$&  $0.85\pm  0.05$ & $7.89\pm  0.28$\\
 $62.44\pm  0.32$&  $3.82\pm  0.82$&  $0.31\pm  0.09$ & $1.26\pm  0.25$\\
 $73.65\pm  0.41$& $11.39\pm  0.97$&  $0.41\pm  0.05$ & $4.92\pm  0.36$\\
 $93.22\pm  0.34$&  $5.67\pm  0.82$&  $0.48\pm  0.10$ & $2.89\pm  0.47$ \\
 $102.31\pm  1.26$&  $9.84\pm  3.23$&  $0.17\pm  0.08$ & $1.81\pm  0.52$ \\
\hline                  
\end{tabular}
\end{table}

\begin{figure*}
\center
\includegraphics[width=0.76\linewidth]{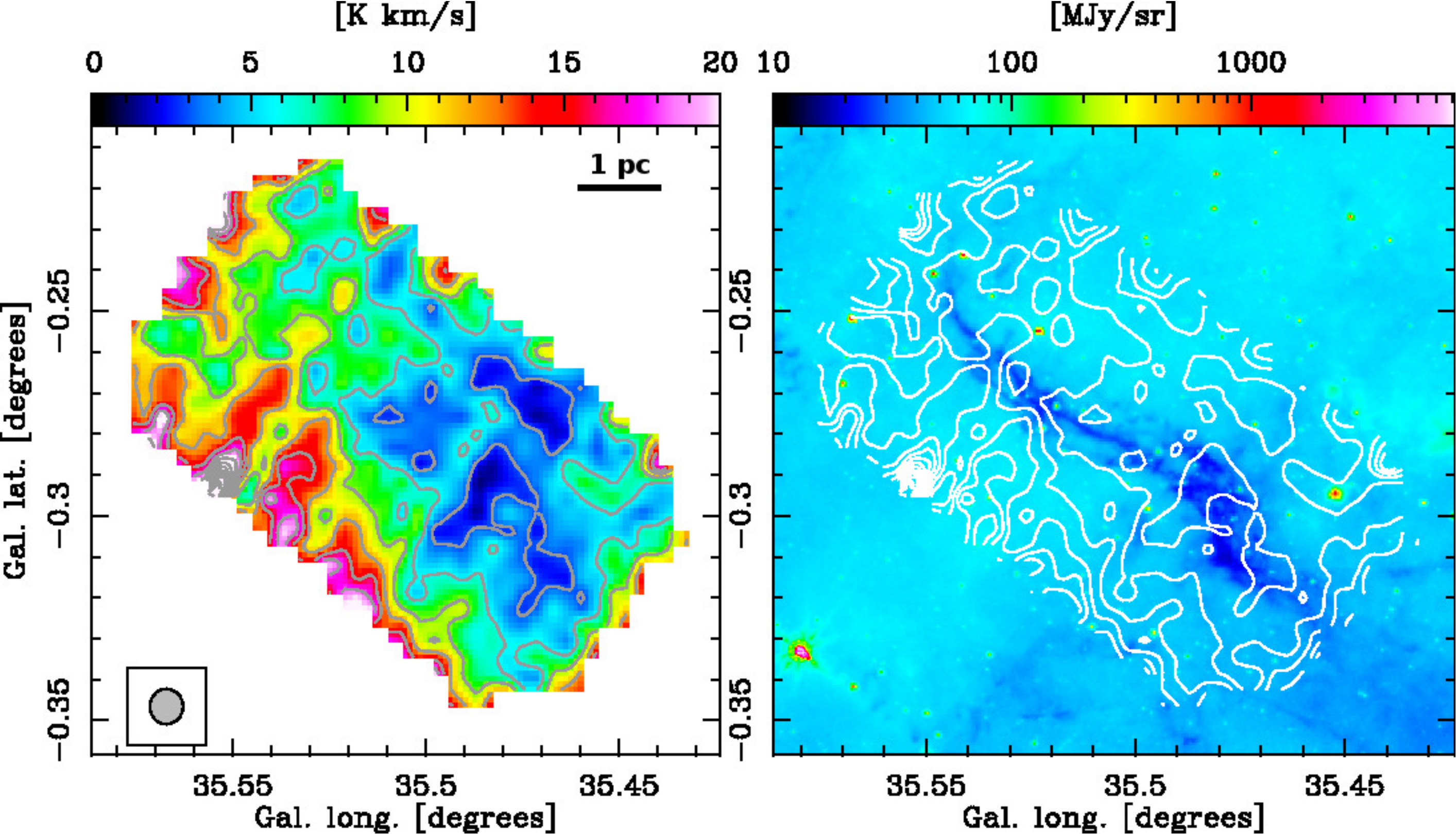}
\caption{
{\it Left:} Integrated emission of [CII] between 40 to
55\,\kms\ corresponding to the brightest velocity component close to
the \vlsr\ of IRDC H. To achieve a better signal-to-noise, we smoothed
the map to 31\arcsec. The beam is shown in the lower left panel. A
linear scale of 1~pc for an adopted distance of 2.9~kpc is shown. {\it
  Right:} {\it Spitzer} 8\,$\mu$m map of the region. IRDC H is
seen in absorption against the Galactic background. White contours
show the velocity integrated \CII\ emission and are the same as in the
left panel.}\label{fig:cii_map}
\end{figure*}

Figure~\ref{fig:cii_map} shows the integrated emission of \CII\ (in
the $40-55\,{\rm km}\,{\rm s}^{-1}$ velocity range) and $8\mu$m
emission map of that area. IRDC~H is seen as a dark filamentary structure,
indicating high extinction and therefore high column densities with
respect to the surrounding medium. 
These maps reveal a spatial offset between the \CII\ emitting gas and
the highest column density regions of the IRDC.
The other components of the spectra are not well detected in the
individual beams, suggesting that they
originate from a more diffuse Galactic foreground or background.

We calculate the column density of C{\sc ii} in the optically thin
limit for the component associated with the IRDC.  Following
\citet{Gold12} \citep[see also][]{Beut14}, the column density,
$N$(C{\sc ii}), is given by the expression
\begin{align}
N({\rm CII})=&\frac{10^{16}}{3.43}\left[1+0.5e^{91.25/T_{\rm kin}}\left(1+\frac{2.4\times10^{-6}}{C_{ul}}\right)\right]\nonumber\\
&\times\int T_{\rm mb}dv,
\end{align}
where $T_{\rm kin}$ is the gas temperature and $C_{ul}$ is the
collision rate.
The integrated emission of the spectrum of the main component is 
$\int T_{\rm mb}dv\sim7.89\,{\rm K}\,{\rm km}\,{\rm s}^{-1}$ (see Table~\ref{table:3}). 
$C_{ul}$ depends primarily on the collisions with H and H$_2$.  
For an average gas temperature of $T_{\rm kin}\sim 160\,{\rm K}$ 
(a reasonable value for photo-dissociation region (PDR) models, and being 
close to the average value we derive in our theoretical modeling below), and 
an H-nucleus number density in the range of $10^2-10^3\,{\rm cm}^{-3}$, we obtain
a column density in the order of $N({\rm CII})\sim10^{17}-10^{18}\,{\rm cm^{-2}}$.

PDR modeling shows that \CII\ is primarily
emitted from lower density regions when they interact with FUV
radiation from nearby stars \citep[e.g.,][]{Roel07}. In regions of
higher extinction, such as the densest part in the right panel of
Fig.~\ref{fig:cii_map}, the FUV radiation is severely extinguished and
thus \CII\ diminishes, unless very strong shocks or very high
cosmic-ray ionization rates are present, \citep[e.g.,][]{Meij11,
  Bisb15, Bisb17a}. However, these are not expected in the case of IRDC~H,
given its location in the Galaxy.

\begin{figure*}
\centering
\includegraphics[width=0.7\linewidth]{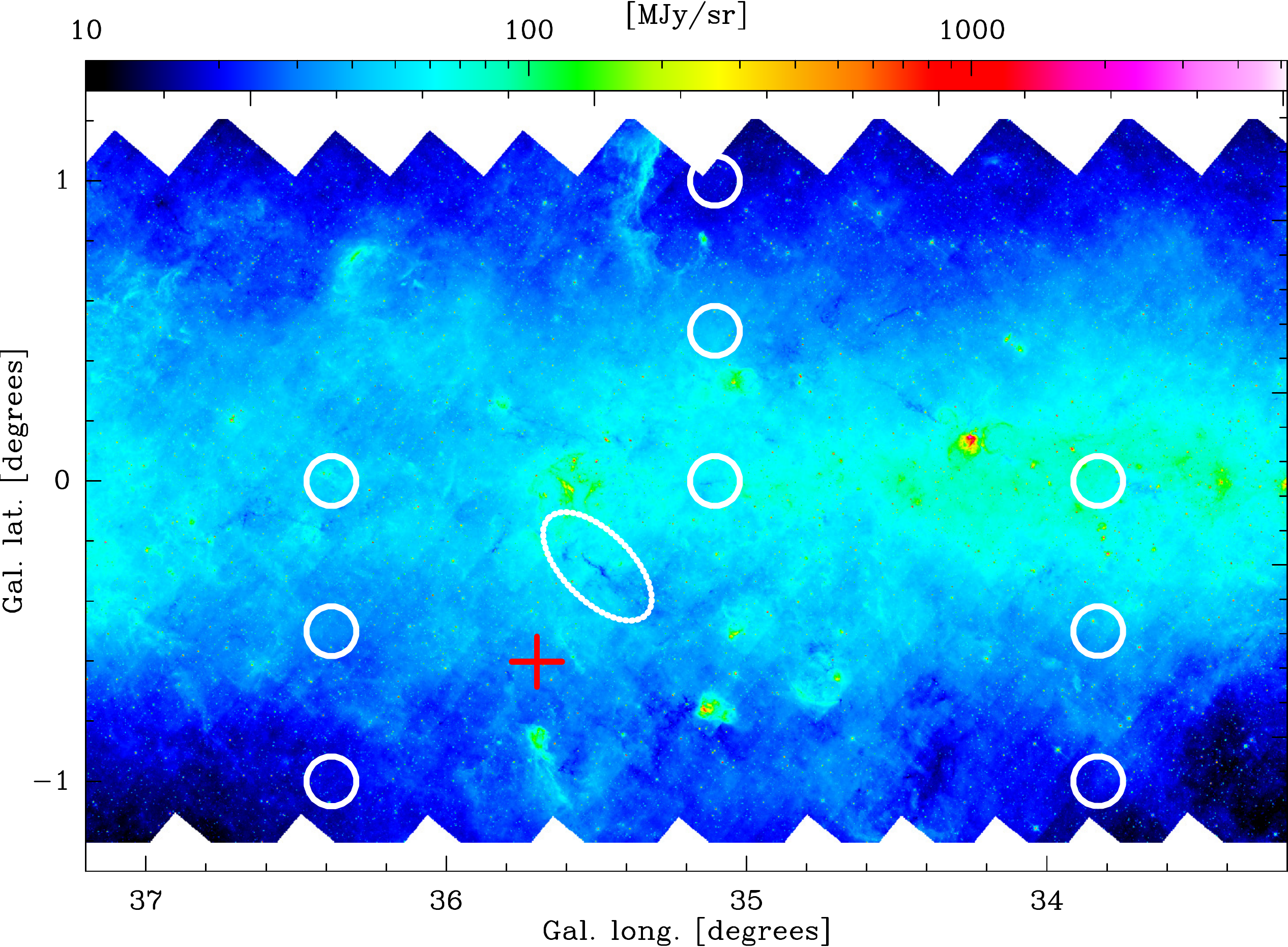}
\includegraphics[width=0.31\linewidth]{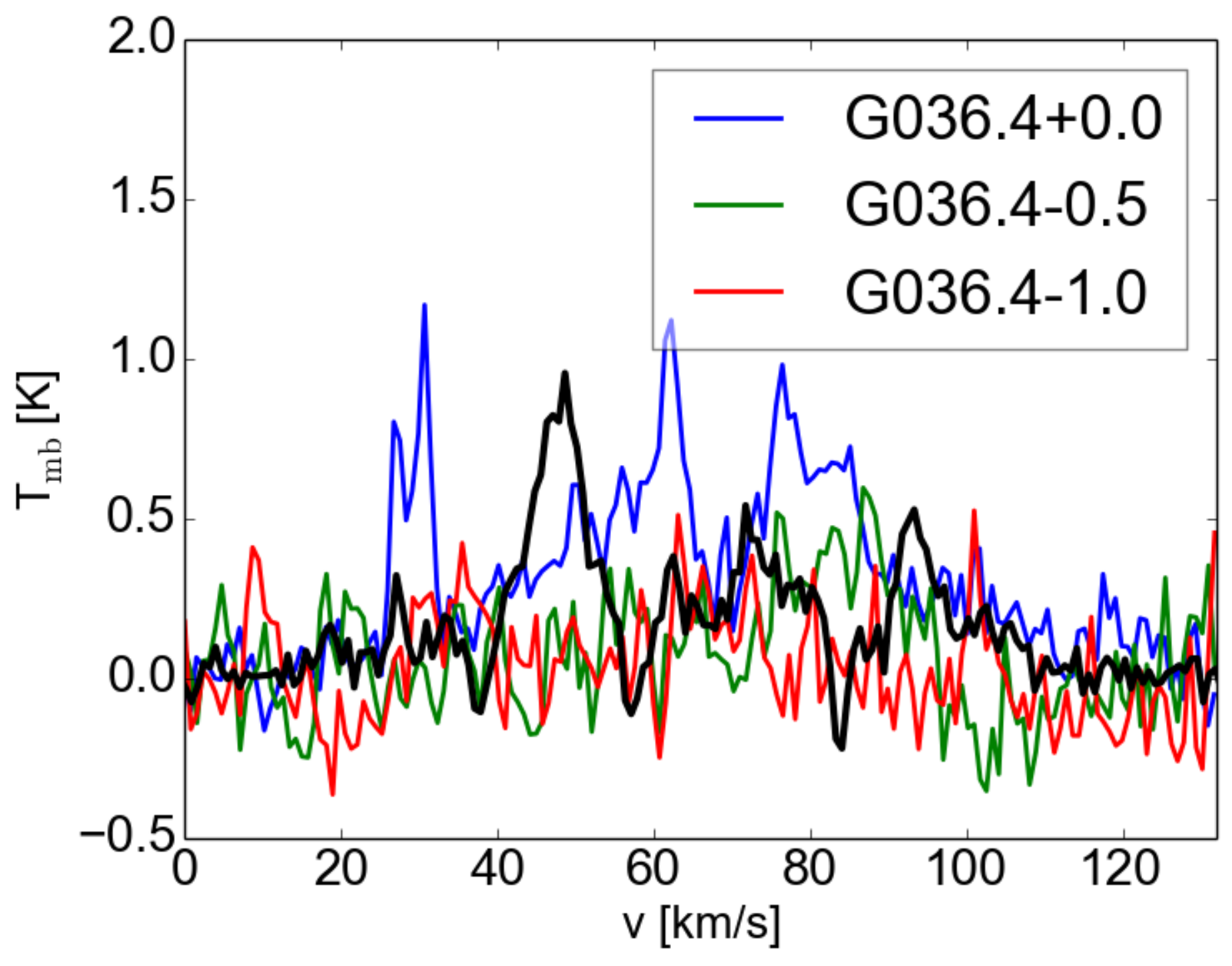}
\includegraphics[width=0.31\linewidth]{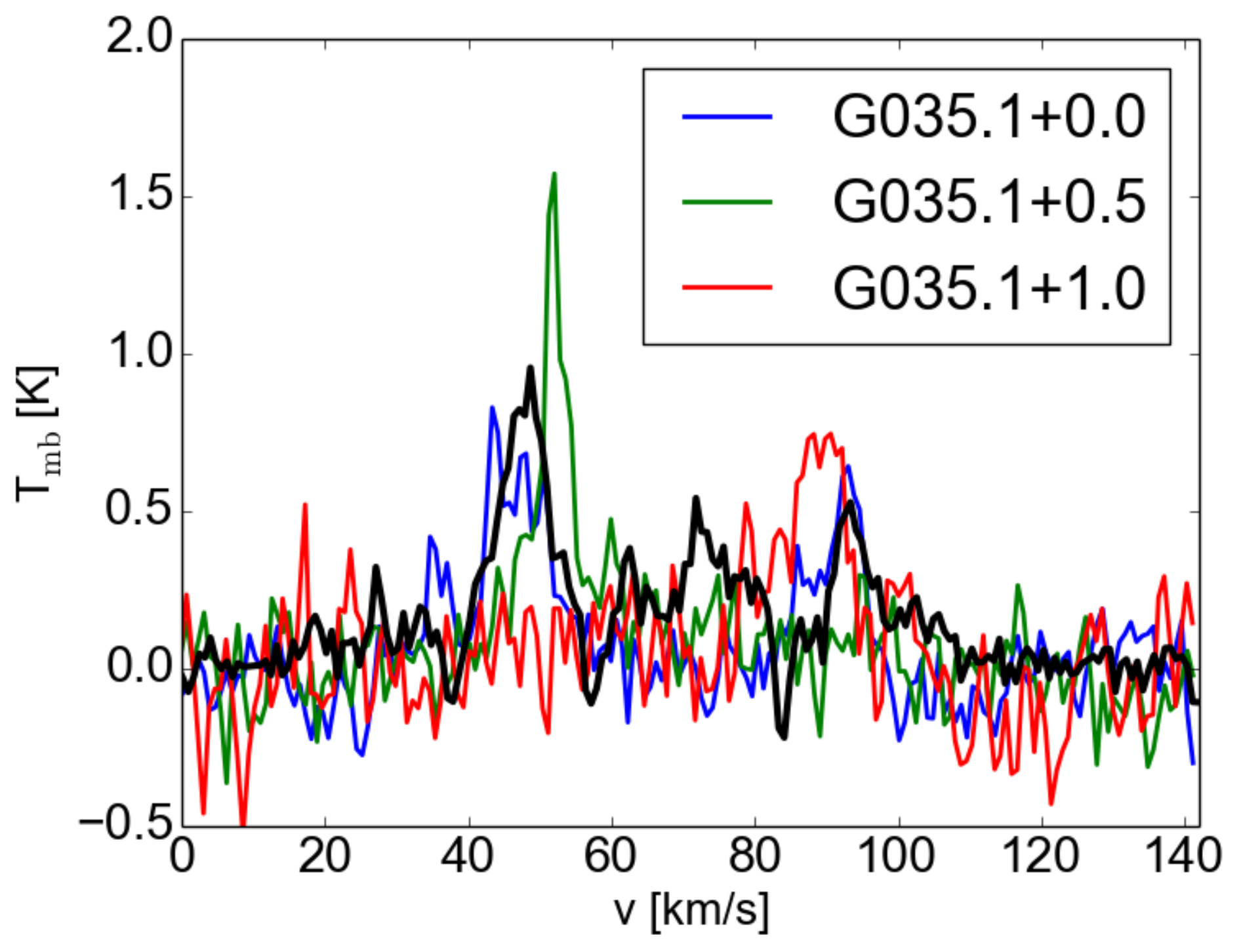}
\includegraphics[width=0.31\linewidth]{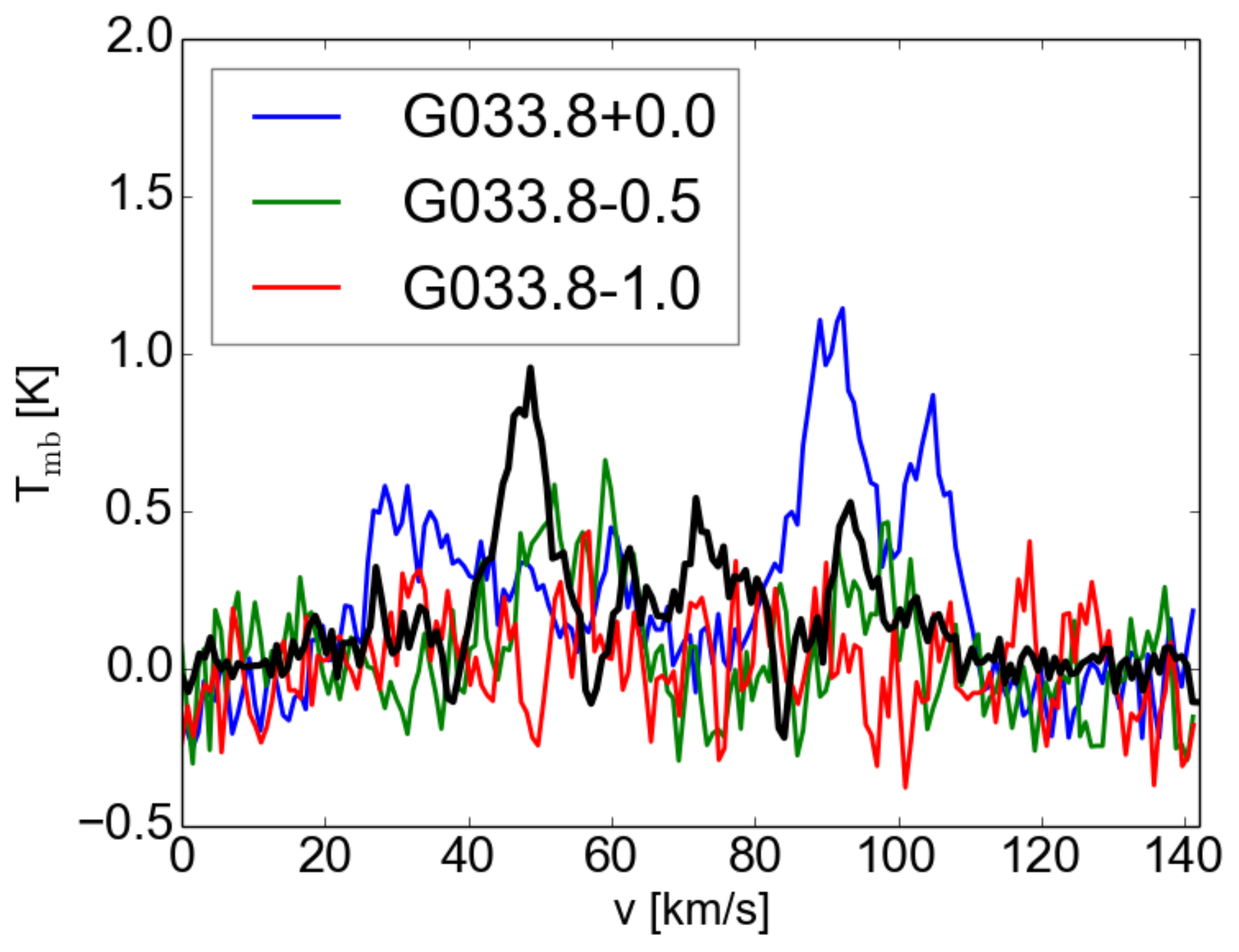}
\caption{
{\it Top panel:} An extended map at $8\mu$m showing the area
surrounding IRDC~H in the Galactic plane. The centre of each white circles is {\it
  Herschel's} \CII\ observations \citep[GOT C+][]{Pine13,Lang14}.
The white ellipse 
indicates the position of IRDC~H, which is close to
the G035+0.0 GOT C+ observation. The red cross shows the OFF
position of our observations. {\it Bottom row:} Blue, green and red
solid lines correspond to the GOT C+ observations of [C{\sc ii}]
$158\,\mu$m for the 9 nearby regions to IRDC~H. The solid black line
in each panel shows the {\it SOFIA} \CII data towards IRDC~H.}
\label{fig:gotc}
\end{figure*}

We compare our IRDC~H data with GOT
C+ data from the {\it Herschel} archive \citep{Pine13,Lang14} of 9
surrounding regions (see Fig.~\ref{fig:gotc}). Assuming a
$\sim2.9\,{\rm kpc}$ distance, i.e., the near kinematic implied by the
velocity of IRDC~H, we find that the closest {\it Herschel}
observation is located $\sim22\,{\rm pc}$ away
(G035.1+0.0). Interestingly, 
%as we see in Fig.~\ref{fig:gotc}, 
$T_{\rm mb}$ peaks at $\sim45\,{\rm km}\,{\rm s}^{-1}$. This indicates a
possible larger-scale connection, meaning that the GMC atomic envelope
extends in that direction, which is in agreement with the
$^{13}$CO(1-0) data analyzed by \citet{Hern15} for that wider
region. This $T_{\rm mb}$ peak is also seen at the G035.1+0.5 {\it
  Herschel} observation, however it is not possible to unambiguously
demonstrate a connection of IRDC~H to that location.
The velocity peak at $v_{\rm lsr}\sim90\,{\rm km}\,{\rm s}^{-1}$ that
we obtain from our data is possibly connected with the kinematics of
the spiral structure of our Galaxy. This is supported from similar
high velocities that are seen in GOT C+ data and in particular in the
objects with Galactic latitude of $-0.5\lesssim b\lesssim0.5$,
contrary to those in higher latitude (see the lower panels of
Fig.~\ref{fig:gotc}).

\section{Theoretical Modeling}

%To explore potential implications of the \CII\ observations for formation scenarios of IRDCs, 
We perform a {\sc 3d-pdr} \citep{Bisb12} modeling of 
subregions extracted from magnetohydrodynamic (MHD) simulations of
GMCs \citep{Wu17a}. We note that we are not attempting to model the specific
structure of IRDC~H, but to understand
the nature of \CII\ emission from similar dense structures that form
via cloud collisions or decay of turbulence in self-gravitating clouds.
We consider two cases: (i) a GMC-GMC collision at a relative speed of 10~$\rm
km\:s^{-1}$ (Model 1 of Wu et al. at $t=4\,{\rm Myr}$); (ii) two overlapping
but non-colliding GMCs along the line-of-sight.
The selected subregions from the MHD simulations have volumes of
$12^3\,{\rm pc}^3$ and, for the case of the colliding GMCs, it contains 
the highest density part of the cloud,
i.e., formed as a direct result of the collision. In this first case, the high density
structure has a linear size of $\lesssim7\,{\rm pc}$, comparable to
the size of IRDC~H ($\sim5\,{\rm pc}$).  We orient it so that the
direction of collision is along the line-of-sight, which maximizes
\vlsr\ offsets that result from the collision \citep{Bisb17}.  
After some experimentation involving comparing total integrated intensities
of \CII\ emission with the observed region, for both the simulations shown
we used an isotropic FUV radiation field of $\chi/\chi_0=10$
\citep[normalized relative to][]{Drai78} and a cosmic-ray ionization
rate of $\zeta_{\rm CR}=10^{-16}\,{\rm s}^{-1}$. This few-times-enhanced FUV intensity
\citep{Wolf03} may indicate possible nearby star-formaion activity, which
is also consistent with relatively bright 8~$\rm\mu$m emission in the surroundings (Fig.~\ref{fig:gotc}).

\begin{figure}
\includegraphics[width=0.48\linewidth]{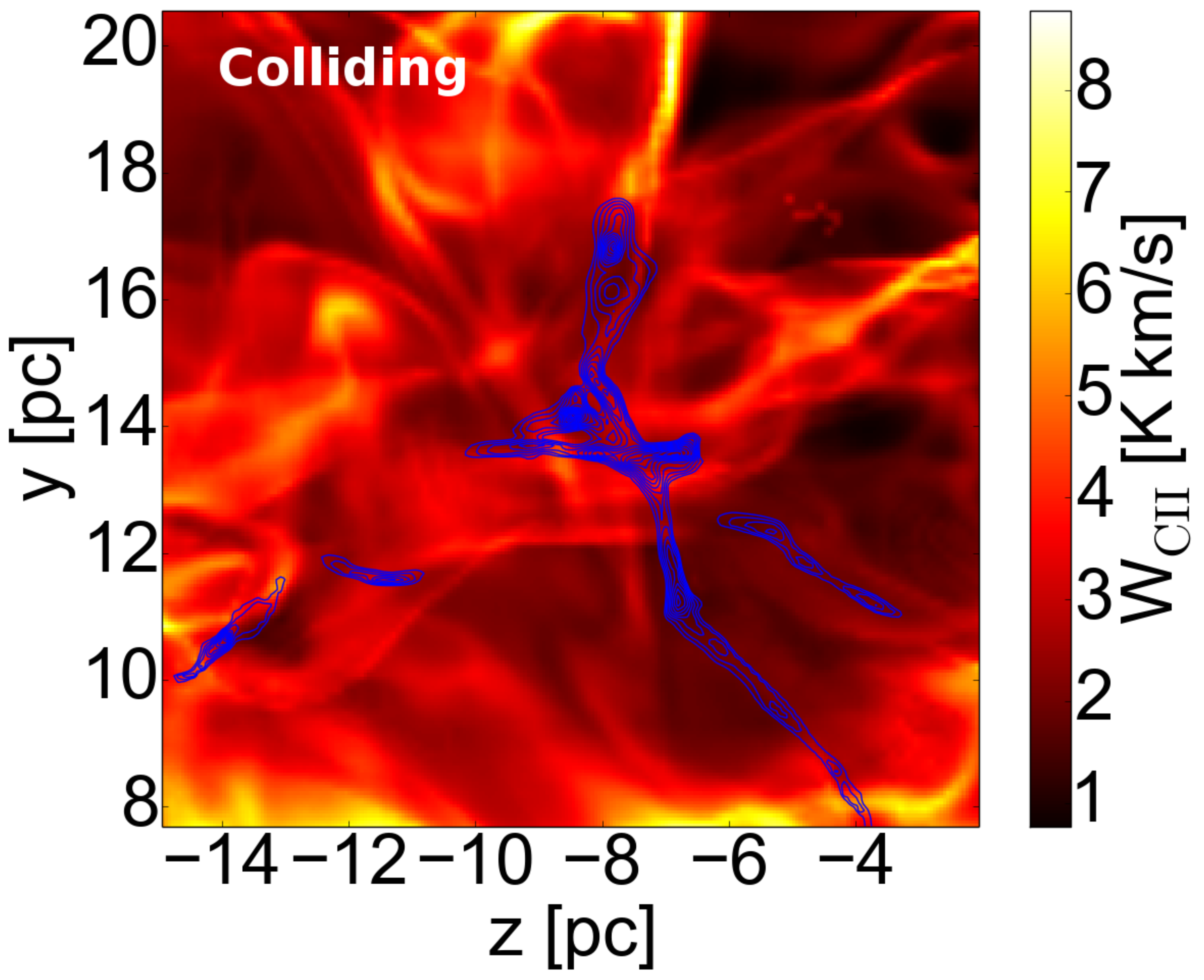}
\includegraphics[width=0.48\linewidth]{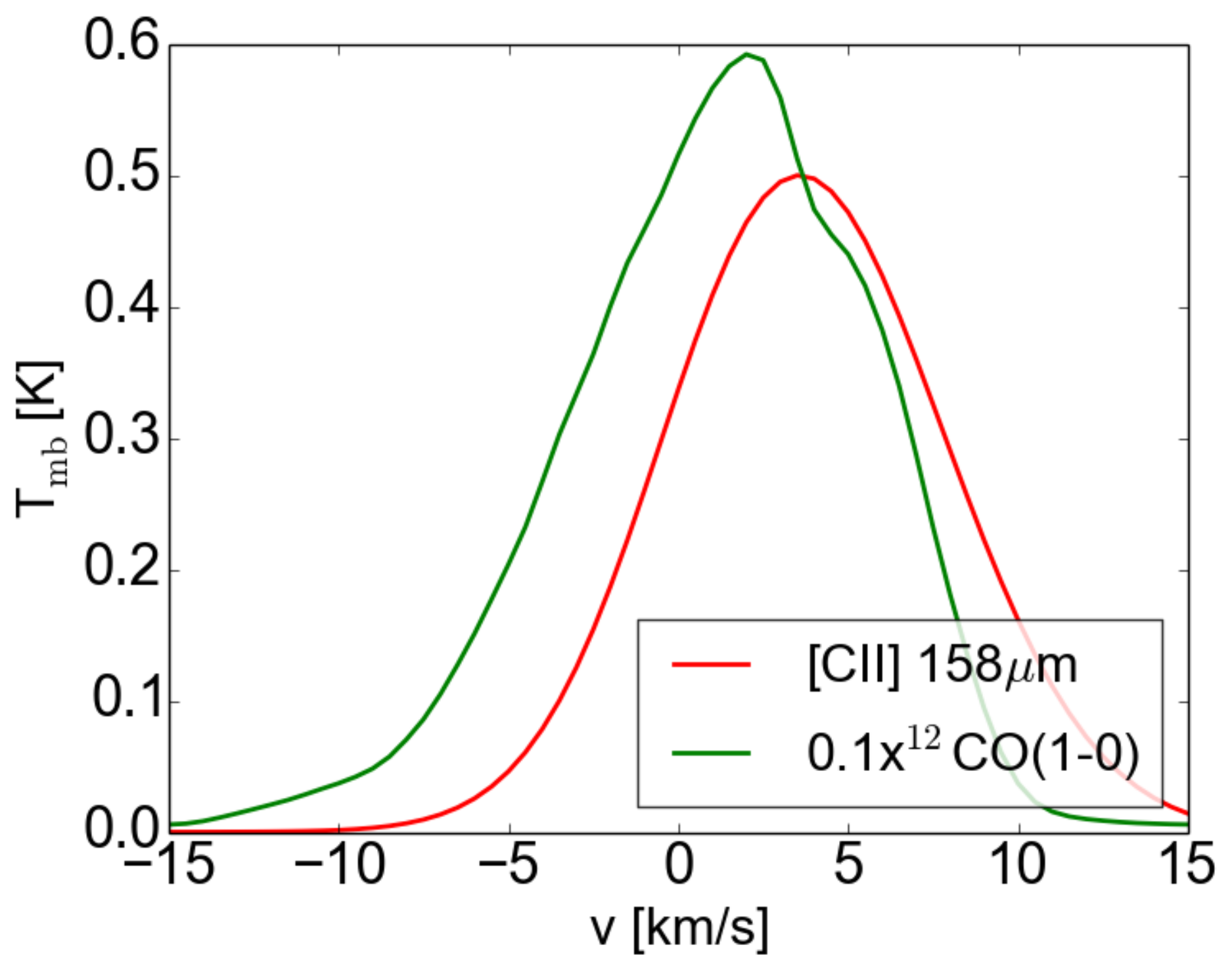}
\includegraphics[width=0.5\linewidth]{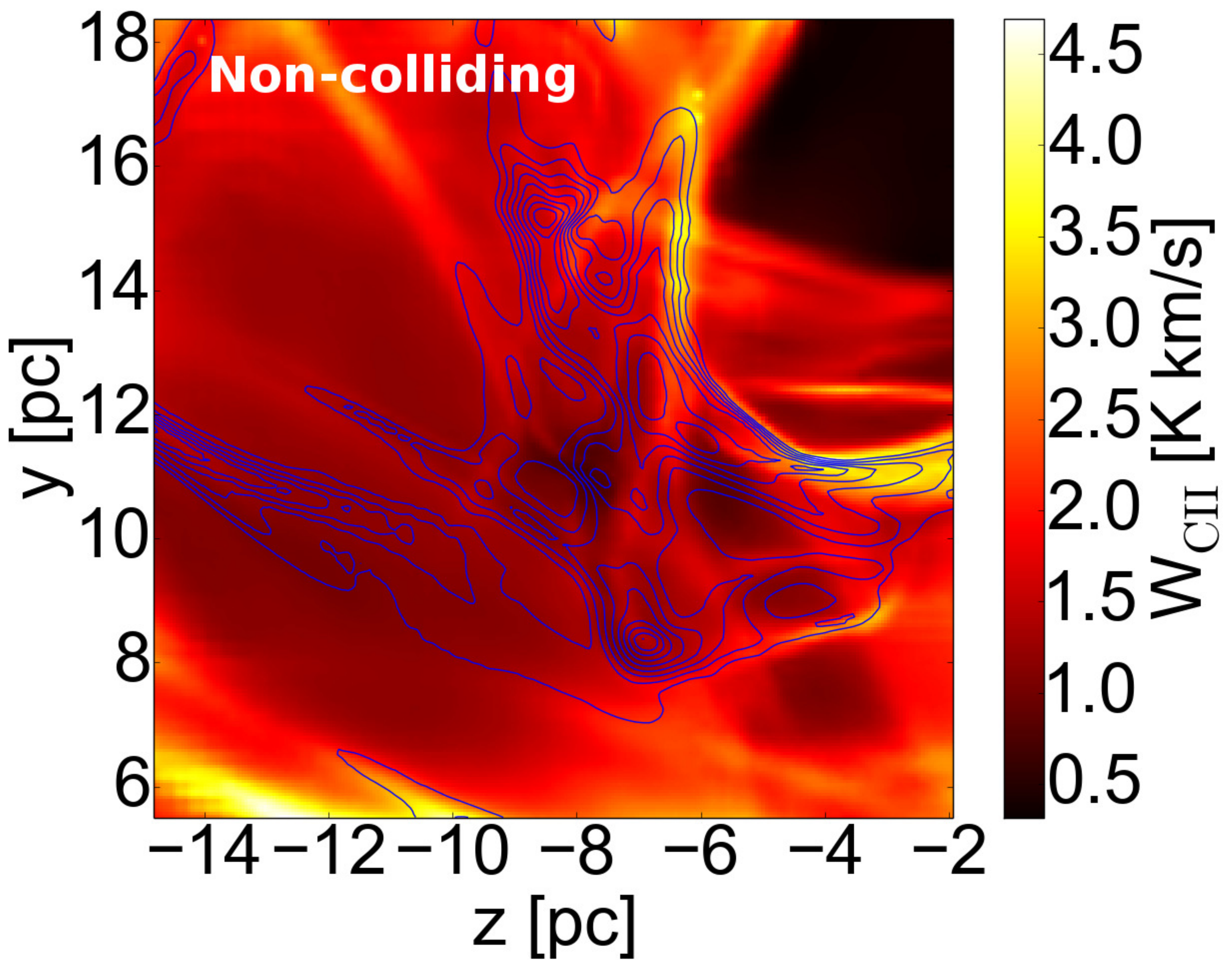}
\includegraphics[width=0.5\linewidth]{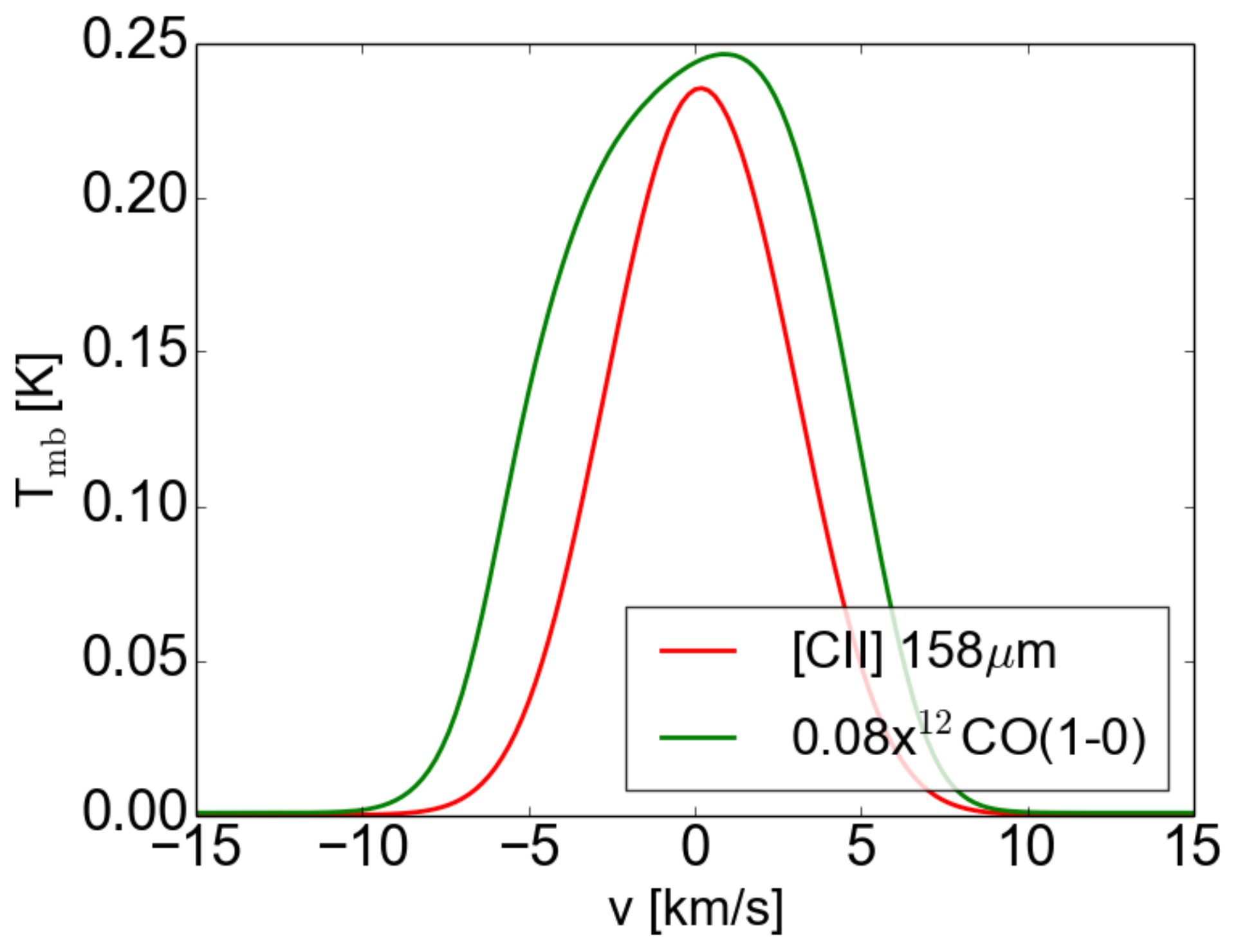}
\caption{
Synthetic observations of a sub-region from the MHD simulation by
\citet{Wu17a} of two colliding (top row) and non-colliding (bottom row) GMCs
that are along the line-of-sight of the observer. The left
column shows the \CII\ emission whereas the blue contours indicate the
highest peaks of the total H-nucleus column density. In the colliding
case, there is an offset of a few pc between [C{\sc ii}] emission and the
column density peaks, in qualitative agreement with our observations of IRDC~H.
In the non-colliding case, we do not find such an offset. The right column
shows the average velocity spectrum of \CII\ and $^{12}$CO $J=1-0$ lines. In the 
colliding case, we find an offset in the $T_{\rm mb}$-weighted average velocities 
of these two lines of about $\rm 2.7\:km\:s^{-1}$. In the non-colliding case,
this offset is $\lesssim0.4\:{\rm km}\:{\rm s}^{-1}$.}
\label{fig:sim}
\end{figure}

The left column of Fig.~\ref{fig:sim} shows the integrated intensity
map of [C{\sc ii}] for the colliding scenario (top panel), and the
non-colliding scenario (bottom panel), without considering any observational
  or instrumental effects.  Contours show the highest peaks
of total H-nucleus column density, $N_{\rm H}$. We find that these are
offset when compared to the peaks of [C{\sc ii}] emission only in the GMC-GMC collision
simulation. This is in qualitative agreement with our observations, indicating that
the observed [C{\sc ii}] originates from the surrounding medium and
not from the dense filament itself.

The right column of Fig.~\ref{fig:sim} shows the average velocity
spectrum of the entire analyzed region of the clouds (top panel: colliding scenario;
bottom panel: non-colliding scenario). The red line corresponds to
the \CII\ emission and the green line to the $^{12}$CO $J=1-0$
emission (we note that $^{13}$CO is not yet treated in the {\sc
  3d-pdr} code, so we use $^{12}$CO as its proxy). In the first case, we find a
  velocity offset of $2.7\,{\rm km}\,{\rm s}^{-1}$ between the average ($T_{\rm mb}$-weighted)
  velocities of \CII\ and $^{12}$CO and about a $1.1\,{\rm km}\,{\rm
    s}^{-1}$ offset between the peaks of these spectra. These offsets
  are similar to those observed between \CII\ and $^{13}$CO towards
  IRDC H. 
The range of the \CII\ velocity distribution is about $10\,{\rm
  km}\,{\rm s}^{-1}$ with a dispersion of $4.1\:{\rm km\:s}^{-1}$.   
The $^{12}$CO emission is distributed over a similar range of
velocities.
Note that $^{13}$CO is expected to be more localized in velocity
space, given that it traces denser conditions. 
The velocity offset and dispersion of the \CII\ and $^{13}$CO in this
theoretical model are broadly consistent with those seen in IRDC~H. 
On the other hand, the non-colliding
  case has a much smaller velocity offset, i.e., the $T_{\rm mb}$-weighted average velocities of \CII\ and $^{12}$CO are both $\lesssim0.4\,{\rm km}\,{\rm s}^{-1}$ and the offset between the peaks of these
spectra is $\sim0.5\,{\rm km}\,{\rm s}^{-1}$. These results thus support the scenario that IRDC~H is a result of the collision between GMCs.

%%%%%%%%%%%%%%%%%%%%%%
\section{Discussion and Conclusions}%
\label{sec:con}      %
%%%%%%%%%%%%%%%%%%%%%%

We have presented {\it SOFIA-upGREAT} observations of [C{\sc ii}]
$158\mu$m emission of a well-studied filamentary IRDC. We find that
the integrated emission of [C{\sc ii}] is spatially offset compared to
the dense structure of the IRDC, as seen in absorption at $8\mu{\rm
  m}$ against the Galactic background. An offset in velocity of about
3~$\rm km\:s^{-1}$ is also seen when comparing [C{\sc ii}] with
$^{13}$CO(1-0) emission. The velocity dispersion of the
\CII\ component is about $9\rm km\:s^{-1}$, a few times broader than
that of the $^{13}$CO(2-1) emission from the IRDC. The column density
of C{\sc ii} of the main component in the optically thin limit
is found to be in the order of $N_{\rm CII}\sim 10^{17}-10^{18}\,{\rm cm}^{-2}$,
assuming an average gas kinetic temperature of $\sim 160\,{\rm K}$, derived 
from our theoretical modeling. A
similar component of \CII\ emission is seen in a nearby region, 22~pc
away to the northwest, from {\it Herschel}-GOTC+ data, but not in the
opposite direction. However, in the local map obtained by {\it SOFIA},
the \CII\ emission is stronger in a direction to the east.

A primary $T_{\rm mb}$ peak in $^{13}$CO(1-0) is observed at
  \vlsr$\sim45\:{\rm km}\:{\rm s}^{-1}$, while a secondary one is
  observed at \vlsr$\sim55\:{\rm km}\:{\rm s}^{-1}$.  Such a
  difference of $\sim10\:{\rm km}\:{\rm s}^{-1}$, 
together with their
  known connection in position-velocity space \citep{Hern15} and the
  observed widespread SiO(2-1) emission \citep{Jime10}, strengthens
  the GMC-GMC collision scenario as the most probable formation
  mechanism of IRDC~H. This is the first such example of a GMC
  collision producing dense gas that likely leads to star cluster
  formation.

Using {\sc 3d-pdr}, we post-processed sub-regions of MHD
simulations of two colliding and non-colliding GMCs, presented in \citet{Wu17b}.
The simulated \CII\ velocity integrated emission maps of the colliding scenario show a spatial
offset when compared to the total H-nucleus column density and are
thus in qualitative agreement with the {\it SOFIA} observations of the
IRDC. Here, a velocity offset is present in the simulated spectra, similar
to the one observed in the IRDC. In the context of a GMC-GMC collision
scenario for the formation of IRDC H, this demonstrates consistency
with the adopted parameters of a 10~$\rm km\:s^{-1}$ collision speed,
which is also supported by the observed $^{13}$CO(1-0) components. These features
are either minimized or not observed in the non-colliding GMCs simulation, supporting
the fact that the IRDC~H may have formed as a result of a cloud-cloud collision.

There are still alternative possibilities to the above. 
For example, the \CII\ emission may simply trace the
turbulent boundary region of a massive GMC or GMC complex, of which
IRDC H only represents one small part, however mapping of a larger region 
around the IRDC is needed. Higher sensitivities are also needed to enable
higher resolution spatial mapping of the \CII\ emission.
Such observations can help improve our understanding of IRDC formation and thus the
processes that initiate star cluster formation and control the star
formation rates of galaxies.

\section*{Acknowledgements}

The authors thank an anonymous referee for providing comments that
improved the clarity of this work. 
TGB and JCT acknowledge support from a grant from
NASA/USRA in support of these observations. TC acknowledges support
from the \emph{Deut\-sche For\-schungs\-ge\-mein\-schaft, DFG\/} via
the SPP (priority programme) 1573 `Physics of the ISM'.


\begin{thebibliography}{}
\bibitem[Barnes et al.(2016)]{Barn16} Barnes, A.~T., Kong, S., Tan, J.~C., et al.\ 2016, \mnras, 458, 1990 
\bibitem[Beuther et al.(2014)]{Beut14} Beuther, H., Ragan, S.~E., Ossenkopf, V., et al.\ 2014, \aap, 571, A53 
\bibitem[Bisbas et al.(2012)]{Bisb12} Bisbas, T.~G., Bell, T.~A., et al.\ 2012, \mnras, 427, 2100
\bibitem[Bisbas et al.(2015)]{Bisb15} Bisbas, T.~G., Papadopoulos, P.~P., \& Viti, S.\ 2015, \apj, 803, 37 
\bibitem[Bisbas et al.(2017a)]{Bisb17a} Bisbas, T.~G., van Dishoeck, E.~F., et al.\ 2017a, \apj, 839, 90 
\bibitem[Bisbas et al.(2017b)]{Bisb17} Bisbas, T.~G., Tanaka, K.~E.~I., et al.\ 2017b, \apj, 850, 23 
\bibitem[Butler \& Tan(2009)]{Butl09} Butler, M.~J., \& Tan, J.~C.\ 2009, \apj, 696, 484 
\bibitem[Butler \& Tan(2012)]{Butl12} Butler, M.~J., \& Tan, J.~C.\ 2012, \apj, 754, 5 
\bibitem[Draine(1978)]{Drai78} Draine, B.~T.\ 1978, \apjs, 36, 595 
\bibitem[Egan et al.(1998)]{Egan98} Egan, M.~P., Shipman, R.~F., Price, S.~D., et al.\ 1998, \apjl, 494, L199 
\bibitem[Goldsmith et al.(2012)]{Gold12} Goldsmith, P.~F., Langer, W.~D., et al.\ 2012, \apjs, 203, 13 
\bibitem[Guan et al.(2012)]{Guan2012} Guan, X., Stutzki, J., Graf, U.~U., et al.\ 2012, \aap, 542, L4 
\bibitem[Hacar et al.(2013)]{Haca13} Hacar, A., Tafalla, M., Kauffmann, J., \& Kov{\'a}cs, A.\ 2013, \aap, 554, A55 
\bibitem[Henshaw et al.(2013)]{Hens13} Henshaw, J.~D., Caselli, P., Fontani, F., et al.\ 2013, \mnras, 428, 3425 
\bibitem[Henshaw et al.(2014)]{Hens14} Henshaw, J.~D., Caselli, P., et al.\ 2014, \mnras, 440, 2860 
\bibitem[Hernandez et al.(2011)]{Hern11} Hernandez, A.~K., Tan, J.~C., Caselli, P., et al.\ 2011, \apj, 738, 11 
\bibitem[Hernandez et al.(2012)]{Hern12} Hernandez, A.~K., Tan, J.~C., Kainulainen, J., et al.\ 2012, \apjl, 756, L13 
\bibitem[Hernandez \& Tan(2015)]{Hern15} Hernandez, A.~K., \& Tan, J.~C.\ 2015, \apj, 809, 154 
\bibitem[Jackson et al.(2006)]{Jack2006} Jackson, J.~M., Rathborne, J.~M., Shah, R.~Y., et al.\ 2006, \apjs, 163, 145 
\bibitem[Jim{\'e}nez-Serra et al.(2010)]{Jime10} Jim{\'e}nez-Serra, I., Caselli, P., Tan, J.~C., et al.\ 2010, \mnras, 406, 187 
\bibitem[Kainulainen \& Tan(2013)]{Kain13} Kainulainen, J., \& Tan, J.~C.\ 2013, \aap, 549, A53 
\bibitem[Langer et al.(2014)]{Lang14} Langer, W.~D., Velusamy, T., et al.\ 2014, \aap, 561, A122 
\bibitem[Meijerink et al.(2011)]{Meij11} Meijerink, R., Spaans, M., et al.\ 2011, \aap, 525, A119 
\bibitem[Molinari et al.(2016)]{Moli16} Molinari, S., Schisano, E., Elia, D., et al.\ 2016, \aap, 591, A149 
\bibitem[P\'erault et al.(1996)]{Pera96} P\'erault, M., Omont, A., Simon, G., et al.\ 1996, \aap, 315, L165 
\bibitem[Pineda et al.(2013)]{Pine13} Pineda, J.~L., Langer, W.~D., et al.\ 2013, \aap, 554, A103 
\bibitem[Risacher et al.(2016)]{Risa16} Risacher, C., G{\"u}sten, R., Stutzki, J., et al.\ 2016, \aap, 595, A34 
\bibitem[R{\"o}llig et al.(2007)]{Roel07} R{\"o}llig, M., Abel, N.~P., Bell, T., et al.\ 2007, \aap, 467, 187 
\bibitem[Simon et al.(2006)]{Simo06} Simon, R., Jackson, J.~M., et al.\ 2006, \apj, 639, 227 
\bibitem[Sokolov et al.(2017)]{Soko17} Sokolov, V., Wang, K., Pineda, J.~E., et al.\ 2017, \aap, 606, A133 
\bibitem[Tan et al.(2014)]{Tan14} Tan, J.~C., Beltr{\'a}n, M.~T., Caselli, P., et al.\ 2014, PPVI, 149 
\bibitem[Wolfire et al.(2003)]{Wolf03} Wolfire, M.~G., McKee, et al.\ 2003, \apj, 587, 278 
\bibitem[Wu et al.(2015)]{Wu15} Wu, B., Van Loo, S., Tan, J.~C., \& Bruderer, S.\ 2015, \apj, 811, 56
\bibitem[Wu et al.(2017a)]{Wu17a} Wu, B., Tan, J.~C., Nakamura, F., et al.\ 2017a, \apj, 835, 137
\bibitem[Wu et al.(2017b)]{Wu17b} Wu, B., Tan, J.~C., Christie, D., et al.\ 2017b, \apj, 841, 88
\end{thebibliography}
\end{document}